\documentclass[aps,preprint,showpacs]{revtex4-1}

\usepackage{amsmath,amssymb,bm,graphicx,epsfig,psfrag,ulem,color,natbib}
\usepackage{units}
\usepackage{float} 
\usepackage{amsmath}   
\usepackage{upgreek}  
\usepackage{bm}  
\usepackage[latin1]{inputenc}
\usepackage{epstopdf}
\bibpunct{}{}{,}{s}{}{,}

\begin{document}

\title{Nonlinear dependence observed in quadrupolar collective excitation of a trapped BEC}

\author{A.R.~Fritsch, P.E.S.~Tavares, Y.R.~Tonin, A.~Bahrami, E.A.L.~Henn, V.S.~Bagnato, G.D.~Telles}

\affiliation{Instituto de F\'{\i}sica de S\~{a}o Carlos, Universidade de S\~{a}o Paulo, C.P. 369, 13560-970 S\~{a}o Carlos, SP, Brazil}
\email{gugs@ifsc.usp.br}

\begin{abstract}
   We report the experimental observation of the collective excitations induced in a magnetically trapped $^{87}$Rb Bose-Einstein condensate. Low-lying mode excitations were studied by tracking the condensate's center-of-mass displacement, and its aspect ratio as a function of the hold time in the trap. We were able to partially control the modes onset by modulating the amplitude of the additional field gradient used to excite the BEC. The measured excitation frequencies were found to be in good agreement with the literature. We have also found that the modulation amplitude was able to change the phase of the center-of-mass oscillation. Finally, an interesting, non-linear dependence was observed on the condensate aspect ratio as a function of the perturbing amplitude which induces the quadrupolar mode.
\end{abstract}

\maketitle

\section{Introduction}

Dilute Bose-Einstein condensates (BECs) are produced and probed using atomic physics tools~\cite{Ketterle98,Pitaevskii}, and their connection to condensed-matter physics is most clear when studying collective excitations~\cite{Jin96,Mewes96,Edwards96,Stringari96}, as well as the sound propagation~\cite{Andrews97,SKurn98,Simula05}. Much of our understanding of the nature of sound in quantum fluids comes from studies carried out in liquid helium~\cite{Findlay38,Atkins51,Hall56,Whitney62}. Recently, with gaseous condensates, the researchers have turned to these decades-old theories from condensed matter physics to test their validity thoroughly and very successfully~\cite{Billy08,Roati08,Boninsegni12}. Nowadays, these versatile experimental system are even being used as fast complex Hamiltonian solvers, which may provide ``on demand'' solutions to particular interesting problems in condensed matter physics.

The nature of collective excitations in a homogeneous Bose gas depends on the hierarchy of three length scales: the excitation reduced wavelength, $\lambda_{ex}$; the healing length, $\xi=1/\sqrt{8 \pi a_s n_0}$; and the mean free path for collisions, $\lambda_{mfp}$. In a inhomogeneous Bose trapped gas, the excitations are influenced by an additional length: the typical condensate size, the Thomas-Fermi radii. This gives rise to three different excitation regimes: Hydrodynamic, collision and collisionless. The first observations of the low-lying excitations in gaseous BECs were made at JILA~\cite{Jin96} and at the MIT~\cite{Mewes96} in the zero temperature limit. Soon after, Stringari carried on in depth theoretical studies of the collective modes ~\cite{Stringari96,Stringari98,Dalfovo99} creating the base framework to understand the observations. 

In this work, we investigate the collective excitations of $^{87}$Rb condensates held in a cigar-shaped Quadrupole-Ioffe configuration (QUIC) trap~\cite{QUIC}. The collective excitations of the condensate are produced by an additional, time varying, magnetic field gradient superimposed to a QUIC potential. Besides the expected oscillation of the center of mass, we observed the BEC undergoing shape oscillations. The frequency of the observed collective excitations were measured and compared to the theoretical predictions. We see evidence of a non-linear dependence on the amplitude of the perturbing field gradient, not yet predicted by the theory. We believe that the study of collective modes is relevant because they present the fundamental features of quantum degenerate gases. Furthermore, we would like to better understand the vortex nucleation and the onset of turbulence in perturbed condensates.

In the next sections, we present a brief discussion on the collective modes observed in BECs, followed by a description of the experimental setup used to acquire the data. The results are then presented and discussed based on the current theory for the dispersion relations, where good agreement is found for the measured mode frequencies. Finally, the conclusions are presented.

\section{Collective excitations in gaseous BECs}

    Bose-Einstein condensates were object of intensive studies early since they were first produced in gaseous alkali metal samples. The low-lying collective excitations were theoretically investigated and predicted just one year after the initial experimental BEC observations~\cite{Stringari96,Pethick02}. Back then, Stringari~\cite{Stringari96} was able to derive analytic solutions for both isotropic and anisotropic harmonic potentials in the Thomas-Fermi regime. 

The normal modes of a BEC are classified by quantum numbers $(n,l,m)$, where $n$ is the radial quantum number and $l,m$ stand for the total angular momentum and its axial projection quantum numbers, respectively. For cylindrical symmetry (as in our trap), $m$ is still a good quantum number, but $l$ is not. Thus, the normal modes are a superposition of wave functions with the same $m$. Stringari discussed an anisotropic harmonic potential with axial symmetry and showed that the lowest $m=0$ modes are coupled excitations of the $(0,2,0)$ modes, quadrupolar surface oscillation, and also the $(1,0,0)$ symmetries. For a cigar-shaped condensates the lowest modes were predicted at frequencies~\cite{Stringari96,Dalfovo99,Pethick02}: 
\begin{equation}
    \omega^2 (m=\pm l) = l\omega_r^2,
    \label{eq1}
\end{equation}
\begin{equation}
    \omega^2 (m=\pm l-1) = (l-1)\omega_r^2+\omega_z^2,
    \label{eq2}
\end{equation}
and 
\begin{equation}
    \omega^2 (m=0) = \omega_r^2(2+\frac{3}{2}\lambda^2 \mp \frac{1}{2} \sqrt{9\lambda^4-16\lambda^2+16}).
    \label{eq3}
\end{equation}
In the above equations $\omega_z$, $\omega_r$ are the axial, radial trap frequencies which will be properly defined in the next section, and $\lambda$ is the ratio of those two frequencies known as the asymmetry parameter. In the large anisotropy limit $(\lambda<<1)$, the solutions of Eq.~\ref{eq3} above are: $\sqrt{5/2}\omega_z$ and $2\omega_r$. The first, low-lying, shape oscillations associated with the above dispersion relations are: 1) the cylindrically symmetric (slow) $m=0$ quadrupole mode where the axial ($z-$axis) length oscillates out of phase with the radial lengths; 2) the in-phase (fast), mostly radial, $m=1$ breathing mode; and 3) the non-symmetric radial oscillation $m=2$ quadrupole mode where the two radial lengths oscillate out of phase.

\section{Experimental setup}

	The experimental sequence to produce the BEC runs as follows. First, $^{87}\rm{Rb}$ BECs containing about $1-2 \times 10^{5}$ atoms are formed in the $|F=2,m_{F}=+2>$ hyperfine state, with a small thermal fraction in a cigar-shaped  QUIC magnetic trap. In order to achieve the degeneracy, the atoms undergo forced evaporation for $23\unit{s}$, going from $20\unit{MHz}$ down to about $1.71\unit{MHz}$, following a non analytic curve experimentally determined by optimizing the phase space density and the runaway condition in each step. We typically produce samples with no more than 30 or 35\% of thermal fraction, well into the Thomas-Fermi regime. The measured trapping frequencies are: $\omega_z=2\pi \times 21.1(1)\unit{Hz}$ in the symmetry axis, and $\omega_r=2\pi \times 188.2(3)\unit{Hz}$ in the radial direction. The radial trapping frequency depends on the offset field, $B_0$, at the trap bottom and the  magnetic field gradient in the radial direction, which can be written as:
\begin{equation}
    \omega_r=\sqrt{\frac{m_Fg_F\mu_B}{m}\frac{(B'_r)^2}{B_0}},
    \label{radfreq}
\end{equation}
where $m_F$, $\mu_B$ and $g_F$ are the magnetic quantum number, the Bohr magneton, and Land\'{e} g-factor for the hyperfine state $|F=2,m_F=+2>$, and $m$ is the atomic mass. The axial trapping frequency is determined by the axial curvature $B''$ through $\omega_z=\sqrt{m_Fg_F\mu_BB''/m}$. The values of $B_0$, $B'$, and $B''$ are determined based on the geometric parameters of the QUIC coils and the electric current running through them. For our system, the calculated values are $B_0=1.22\unit{G}$, $B'_z=162\unit{G/cm}$, $B'_r=165\unit{G/cm}$, and $B''_z= 282.8\unit{G/cm^2}$. Correspondingly, the theoretical results of the radial and axial trapping frequencies are $\omega_r=2\pi \times 188.6\unit{Hz}$ and $\omega_z=2\pi \times 21.4\unit{Hz}$, respectively, for a current of $25\unit{A}$ through the coils. Finally, we have $m_F=2$, $g_F=1/2$, and the anisotropy parameter is $\lambda=\frac{\omega_z}{\omega_r}=0.113$.

\begin{figure} [ht!]
\begin{center}
\includegraphics[%
  width=0.75\linewidth,
  keepaspectratio]{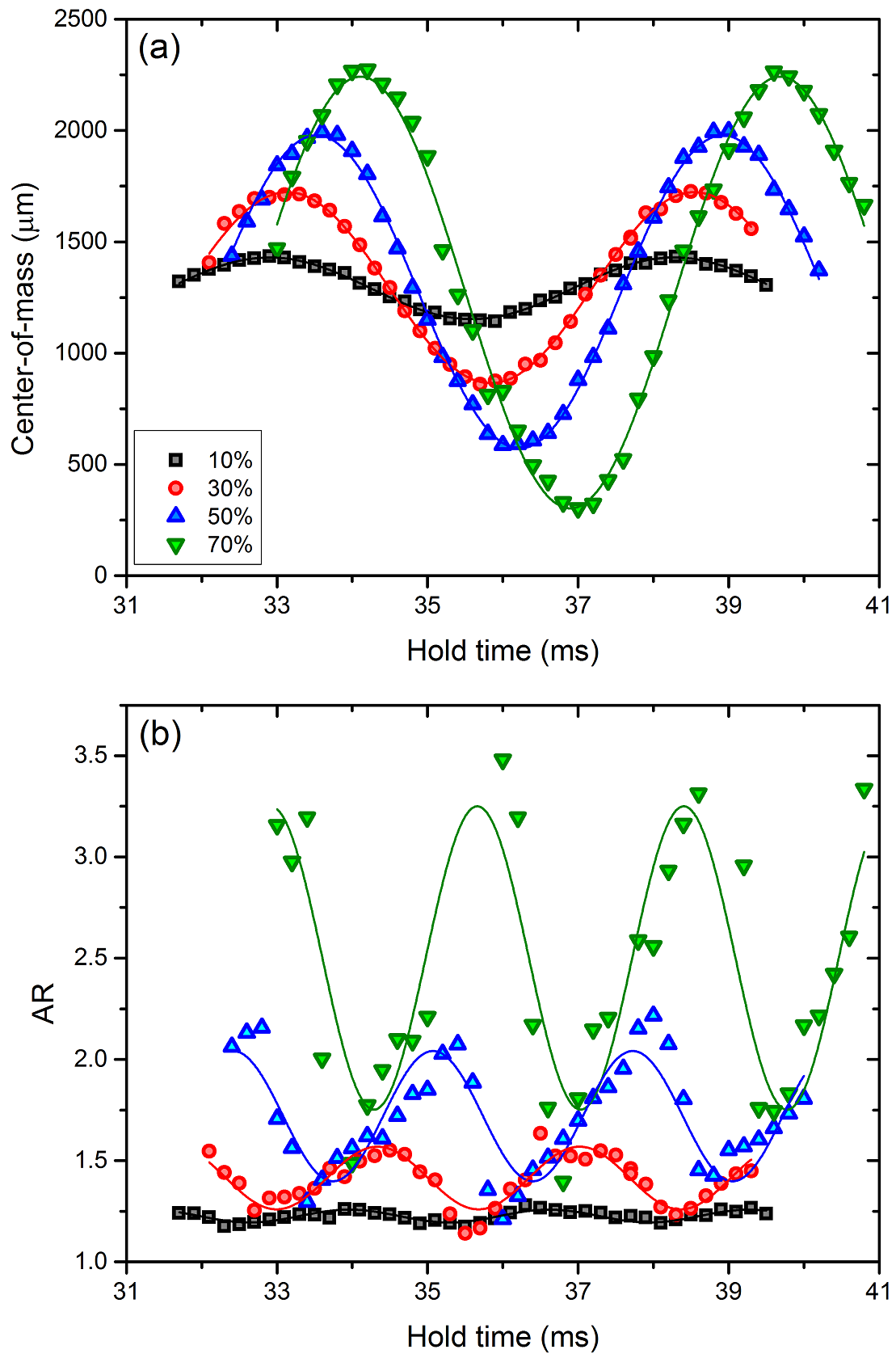}
\end{center}
\caption{Typical dataset acquired during the experimental runs. It shows the collective excitations introduced by the additional magnetic field gradient, for six different hold times, before the trap was switched-off. Please, note the dipolar, quadrupolar, breathing and scissors modes, all coupled, appearing in the normalized absorption image snapshopts taken after $20\unit{ms}$ time-of-flight. The perturbing amplitude increase mostly stretches the BEC radii and displaces its center of mass farther from the rest position (no perturbation).}
\label{fig1}
\end{figure}

To induce the collective excitations, an additional sinusoidal magnetic field gradient is turned on and superimposed on the atoms just after the BEC is produced. The external gradient is a quadrupolar field generated by an anti-Helmholtz coil pair set with symmetry axis closely parallel to the weak trap axis. The coil pair is not perfectly aligned to the condensate axis ($\theta \leq 5^{\circ}$), so its field gradient, near the trapping region, has components parallel to each trap eigen-axis. A superposition of shape morphing, trap bottom displacement and axial rotation is coupled to the BEC. Additional, sinusoidal field gradients going from zero up to $800\unit{mG/cm}$, along the vertical axis, were applied to the trapped atoms inducing the onset of the collective modes. The extra gradient oscillating at $189\unit{Hz}$ was kept on for six extra periods of trap oscillation and then switched-off. In Fig.\ref{fig1}, we present a typical set of the perturbed BEC samples imaged for different hold times, nearly completing a full oscillation cycle.

We then kept the BEC held in the trap for different time intervals, $t_{hold}$, ranging from $30\unit{ms}$ to $50\unit{ms}$ after which we observe it by performing a resonant time-of-flight (TOF) absorption imaging, after $20\unit{ms}$ free fall to avoid pixel saturation and increase the accuracy. Center-of-mass position of the cloud as well as its axes sizes and orientations are recorded and plotted as a function of the $t_{hold}$, from where we extract both frequency and amplitude of the collective modes. 
	
\section{Results and discussions}
	
	Using the technique explained in the previous section, we are able to observe dipolar, quadrupolar, breathing and also the scissors modes in our BECs. For small excitation amplitudes, we observed the tilting about the gravity direction ($y-axis$). Such periodic axial tilting is generated by a sudden rotation of the confining trap, and it is known in the literature as scissors mode. It was theoretically studied by Gu\'ery-Odelin~\cite{Odelin99} and others and will not be discussed here. Here, we will focus on the dipolar and the quadrupolar modes only.
	
After being perturbed, the BEC center-of-mass always oscillate in the dipolar mode~\cite{Jin96}, into the trap. This mode corresponds to the full oscillation at the trap frequency in each direction. We have also observed the onset of the quadrupolar mode~\cite{Mewes96}. Since the longitudinal and transverse directions do not oscillate exactly at $90^{\circ}$ out-of-phase, we believe that the higher order modes may also be onset.	Moreover, the condensate's long axis cyclic tilting was observed and priviously reported~\cite{Henn09}, see Fig.\ref{fig1}. The observation of the scissors mode~\cite{Marago00} shows that the perturbing field gradient is able to cause the BEC to rotate.

\begin{figure} [ht!]
\begin{center}
\includegraphics[%
  width=0.75\linewidth,
  keepaspectratio]{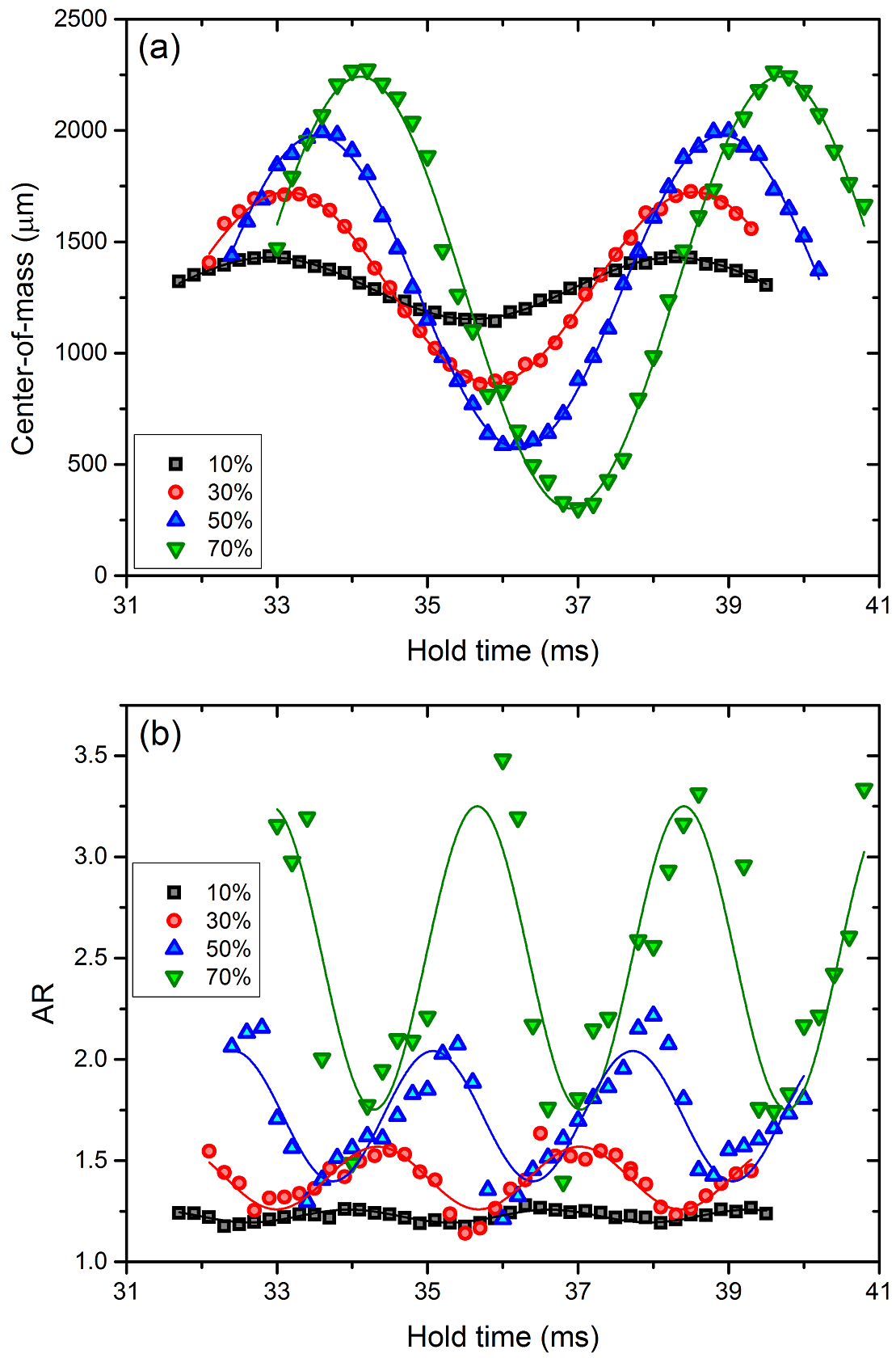}
\end{center}
\caption{(Color online) (a) the center-of-mass (CM) motion as a function of the hold time for different perturbing amplitudes. (b) The AR amplitude versus the hold time, for  4 different amplitudes from 10 to 70\% of $800\unit{mG/cm}$, corresponding to the quadrupolar collective mode.}
\label{fig2}
\end{figure}		
	
	The study of the low-lying collective excitations was carried on by following the center-of-mass (CM) motion and the aspect ratio (AR) of the BECs as a function of the hold time. For small amplitudes, the CM is usually located near the center of the atomic density distribution, which is also the geometric center of the BEC. But, in some situations, at large perturbing amplitudes ($>80\%$),  we observed slosh dynamics taking place, clearly displacing the thermal and condensed fractions in opposite, out-of-phase, CM motion~\cite{Pedro13}. It is important to mention that the dipolar mode preserves the AR when compared to an unperturbed BEC. Therefore, the AR analysis are used only to study the higher order modes (quadrupolar and breathing). 

In Fig.\ref{fig2}(a), we present the results acquired by directly tracking the center-of-mass (CM) motion as a function of the hold time for four different perturbing amplitudes. One may note that, the perturbing energy not only increases the CM displacement from the rest position, but also shifts the phase of the sinusoidal motion. Fig.\ref{fig2}(b) presents the AR evolution with the hold time, measured for the quadrupolar mode. Again, note the phase shift introduced in the AR evolution as the perturbing energy amplitude raises from $80\unit{mG/cm}$ (10\%) to $553\unit{mG/cm}$ (70\%). The data shown in Figs.\ref{fig2}(a) and (b) are fitted by simple sinusoidal functions from where we determine the oscillation frequency of the collective modes and their amplitudes. 

\begin{figure} [ht!]
\begin{center}
\includegraphics[%
  width=0.75\linewidth,
  keepaspectratio]{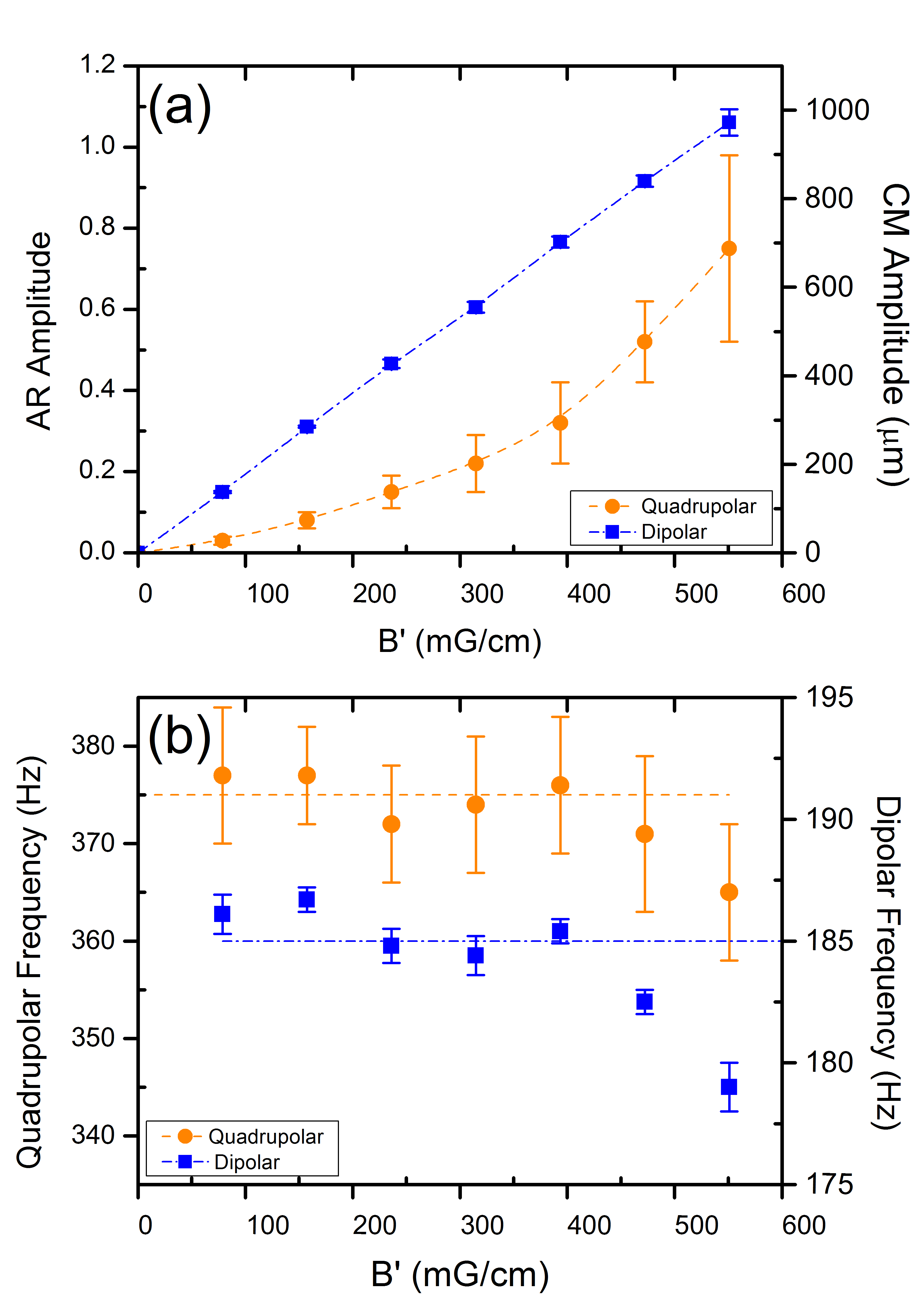}
\end{center}
\caption{(Color online) (a) CM displacement and the AR evolution as a function of the perturbing gradient absolute values. (b) Dipolar and quadrupolar oscillation frequencies versus the perturbing amplitude.}
\label{fig3}
\end{figure}

The evolution of the CM and the AR as a function of the perturbing amplitude is shown in Fig.\ref{fig3}(a). On the one hand, the CM oscillation amplitude proportionally increases with the excitation amplitude starting from zero, as expected. On the other hand, the evolution of the AR clearly show different evolution slopes, depending on perturbing amplitude range. And it displays an interesting non-linear dependence near $400\unit{mG/cm}$. A small region around $350-400\unit{mG/cm}$ presents a interesting non-linear behavior, rapidly evolving from the initial to the final (steeper) slope, strongly suggesting the emergence of a different regime for perturbing amplitudes lager than $400\unit{mG/cm}$. We do not have yet a theoretical explanation for this observation.

Fig.\ref{fig3}(b) presents the changes on the dipolar and quadrupolar oscillation frequencies as the perturbing amplitude increases. We measured $\omega_d=2\pi\times185\unit{Hz}$ for the dipolar oscillations, compared to $\omega_r$ predicted. And,  $\omega_q=2\pi\times375\unit{Hz}=1.99\omega_r$ for the fast oscillation, quadrupolar mode, which is to be compared to the theoretical prediction: $2\omega_r$. The observed modes were previously studied and analytic expressions for the dispersion relations were theoretically derived~\cite{Stringari96,Pethick02}. The agreement found between our results and the predicted frequencies, Eqs.\ref{eq1} to \ref{eq3}, presented in section 2 is very good. The results show almost no change with the frequency for amplitudes up to about $400\unit{mG/cm}$ and a slight decrease ($2-3\%$) for larger amplitudes. Similar observations were seen before~\cite{Mewes96}. We expected to see a constant evolution along the full range and believe the slight change may attributed to nonlinear interactions coupling the condensate normal modes~\cite{Ruprecht95}. For nearly pure condensates, the thermal excitations should have no effect at all and, to the best of our knowledge, there is still no theoretical explanation for a possible slow down of the collective modes frequency of a trapped BEC.
 
We believe though, that the different behavior observed in between the CM oscillation and the shape oscillation (quadrupolar) might be due to way we perturb the BEC. The addition of a small amplitude, time-varying, tilted magnetic field gradient will mostly displace the CM of the BEC. But, once the axial and the radial trapping gradients are a factor of two different, it is enough to induce rotation and shape oscillations. This observation is in agreement with results presented in Figs.\ref{fig1} - \ref{fig3}. Therefore, small perturbation results in CM displacement with little effect on the BEC shape.
 
 As the perturbation amplitude increases, the condensate is pushed farther away from the trap bottom. In a QUIC trap, the BEC needs to be displaced away from the trap center, up to regions with steeper field curvature (see Fig.2(d) in the original paper~\cite{QUIC}), where the unbalanced compression is likely to be effective inducing shape oscillations. That is consistent with the near resonant perturbation used to drive our BEC. Moreover, the QUIC traps are known to be asymmetric along the weak axis direction due to the presence of a single Ioffe coil, making it easier for the condensate to reach the non symmetric boundaries of the trapping potential. This also seems consistent with the observation of a change in the curvature shown in Fig.\ref{fig3}, where the BEC breaks free from the inner harmonic (symmetric) trapping region and travels through the outermost, anharmonic, trap region. These outer trap regions present much stepper field gradient and curvature making them suitable to produce shape oscillations in a BEC by unbalanced compression. Beyond that, we believe that the steep gradient regions might be the inducing mode coupling and and even the dephasing observed in Fig.\ref{fig2}.
 
 This complex coupling dynamics seems particular to our experimental system. It may also be the key to the better understand the vortex nucleation and the turbulence emergence observed in previous experiments~\cite{Seman11}. In contrast, the presented results are in good agreement with the theoretical predictions except for the observed threshold for the quadrupolar shape oscillations.

\section{Conclusions}

	We have studied the excitations induced in cigar-shaped condensates produced in a QUIC trap, near zero temperature. Collective modes were excited by sinusoidally varying currents running through additional coils, externally set for this purpose. The excitation was not axially symmetric, and so the $m\neq0$ modes were excited too. Dipolar oscillations were readily induced and used to accurately measure the trap frequencies. At least, two additional shape oscillations were set on and studied. The higher frequency, primarily, radial breathing mode was excited and its frequency was measured, in very good agreement with the predicted theoretical value. Finally, we have observed an interesting, nonlinear, onset dependence on the minimum excitation amplitude needed to start changing aspect ratio.

\begin{acknowledgements}
We acknowledge financial support from FAPESP, CNPq and CAPES.
\end{acknowledgements}



\begin{thebibliography}{0}%
\makeatletter
\providecommand \@ifxundefined [1]{%
 \@ifx{#1\undefined}
}%
\providecommand \@ifnum [1]{%
 \ifnum #1\expandafter \@firstoftwo
 \else \expandafter \@secondoftwo
 \fi
}%
\providecommand \@ifx [1]{%
 \ifx #1\expandafter \@firstoftwo
 \else \expandafter \@secondoftwo
 \fi
}%
\providecommand \natexlab [1]{#1}%
\providecommand \enquote  [1]{``#1''}%
\providecommand \bibnamefont  [1]{#1}%
\providecommand \bibfnamefont [1]{#1}%
\providecommand \citenamefont [1]{#1}%
\providecommand \href@noop [0]{\@secondoftwo}%
\providecommand \href [0]{\begingroup \@sanitize@url \@href}%
\providecommand \@href[1]{\@@startlink{#1}\@@href}%
\providecommand \@@href[1]{\endgroup#1\@@endlink}%
\providecommand \@sanitize@url [0]{\catcode `\\12\catcode `\$12\catcode
  `\&12\catcode `\#12\catcode `\^12\catcode `\_12\catcode `\%12\relax}%
\providecommand \@@startlink[1]{}%
\providecommand \@@endlink[0]{}%
\providecommand \url  [0]{\begingroup\@sanitize@url \@url }%
\providecommand \@url [1]{\endgroup\@href {#1}{\urlprefix }}%
\providecommand \urlprefix  [0]{URL }%
\providecommand \Eprint [0]{\href }%
\providecommand \doibase [0]{http://dx.doi.org/}%
\providecommand \selectlanguage [0]{\@gobble}%
\providecommand \bibinfo  [0]{\@secondoftwo}%
\providecommand \bibfield  [0]{\@secondoftwo}%
\providecommand \translation [1]{[#1]}%
\providecommand \BibitemOpen [0]{}%
\providecommand \bibitemStop [0]{}%
\providecommand \bibitemNoStop [0]{.\EOS\space}%
\providecommand \EOS [0]{\spacefactor3000\relax}%
\providecommand \BibitemShut  [1]{\csname bibitem#1\endcsname}%
\let\auto@bib@innerbib\@empty
\end{thebibliography}%


\begin{thebibliography}{10}

\bibitem{Ketterle98}
W. Ketterle , D. S. Durfee and D. M. Stamper-Kurn, Making, Probing and Understanding Bose-Einstein Condensates (Varenna: International School of Physics, (1998).

\bibitem{Pitaevskii}
L. Pitaevskii, and S. Stringari, Bose-Einstein Condensation, {\it Oxford University Press} (2003).

\bibitem{Jin96}
D. S. Jin, J. R. Ensher, M. R. Matthews, C. E. Wieman,  and E. A. Cornell, {\it Phys. Rev. Lett.} \textbf{77}, 420, (1996).

\bibitem{Mewes96}
M.-O. Mewes, M. R. Andrews, N. J. van Druten, D. M. Kurn, D. S. Durfee, C. G. Townsend, and W. Ketterle, {\it Phys. Rev. Lett.} \textbf{77}, 988, (1996).

\bibitem{Edwards96}
M. Edwards, P. A. Ruprecht, K. Burnett, R. J. Dodd, and C. W. Clark, {\it Phys. Rev. Lett.} \textbf{77}, 1671, (1996).

\bibitem{Stringari96}
S. Stringari, {\it Phys. Rev. Lett.} \textbf{77}, 2360, (1996).

\bibitem{Andrews97}
M. R. Andrews, D. M. Kurn, H.-J. Miesner, D. S. Durfee, C. G. Townsend, S. Inouye, and W. Ketterle, {\it Phys. Rev. Lett.} \textbf{79}, 553 , (1997).

\bibitem{SKurn98}
D. M. Stamper-Kurn, H.-J. Miesner, S. Inouye, M. R. Andrews, and W. Ketterle, {\it Phys. Rev. Lett.} \textbf{81}, 500, (1998).

\bibitem{Simula05}
T. P. Simula, P. Engels, I. Coddington, V. Schweikhard, E. A. Cornell, and R. J. Ballagh, {\it Phys. Rev. Lett.} \textbf{94}, 080404, (2005).

\bibitem{Findlay38}
J. C. Findlay, A. Pitt, H. G. Smith And J. O. Wilhelm,  {\it Phys. Rev.} \textbf{54}, 506, (1938).

\bibitem{Atkins51}
K. R. Atkins, and C. E. Chase, {\it Proc. Phys. Soc. A} \textbf{64}, 826, (1951).

\bibitem{Hall56}
H. E. Hall, and W. F. Vinen {\it  Proc. R. Soc. Lond. A} \textbf{238}, 204, (1956).

\bibitem{Whitney62}
W. M. Whitney, and C. E. Chase, {\it Phys. Rev. Lett.} \textbf{9}, 243, (1962).

\bibitem{Billy08}
J. Billy, V. Josse, Z. Zuo, A. Bernard, B. Hambrecht, P. Lugan, D. Cl\'ement, L. Sanchez-Palencia, P. Bouyer, and A. Aspect, {\it Nature (London)} \textbf{453}, 891, (2008).

\bibitem{Roati08}
G. Roati, C. D'Errico, L. Fallani, M. Fattori, C. Fort, M. Zaccanti, G. Modugno, M. Modugno, and M. Inguscio, {\it Nature (London)} \textbf{453}, 895, (2008).

\bibitem{Boninsegni12}
M. Boninsegni, and N. V. Prokof'ev, {\it Rev. Mod. Phys.} \textbf{84}, 759, (2012).

\bibitem{Stringari98}
F. Zambelli and S. Stringari, {\it Phys. Rev. Lett.} \textbf{81}, 1754, (1998).

\bibitem{Dalfovo99}
F. Dalfovo, S. Giorgini, L. Pitaevskii, and S. Stringari, {\it Rev. Mod. Phys.} \textbf{71}, 463, (1999).

\bibitem{QUIC}
T. Esslinger, I. Bloch, and T. W. H\"{a}nsch, {\it Phys. Rev. A.} \textbf{58}, R2664, (1998).

\bibitem{Pethick02}
C.J. Pethick and H. Smith , Bose-Einstein Condensation in Dilute Gases (Cambridge: Cambridge University, 2002).

\bibitem{Odelin99}
D. Gu\'{e}ry-Odelin, and S. Stringari, {\it Phys. Rev. Lett.} \textbf{83}, 4452, (1999).

\bibitem{Henn09}
E. A. L. Henn, J. A. Seman, E. R. F. Ramos, M. A. Caracanhas, P. Castilho, E. P. Ol\'{\i}mpio,  G. Roati, D. V Magalh\~{a}es, K. M. F. Magalh\~{a}es, V. S. and Bagnato, {\it Phys. Rev. A} \textbf{79}, 043618, (2009).

\bibitem{Marago00}
O. M. Marag\`{o}, S. A. Hopkins,  J. Arlt, E. Hodby, G. Hechenblaikner, C. J. Foot, {\it Phys. Rev. Lett.} \textbf{84}, 2056, (2000).

\bibitem{Pedro13}
P.E.S. Tavares, G.D. Telles, R.F. Shiozaki, C. Castelo Branco, K.M. Farias, and V.S. Bagnato, {\it Las. Phys. Lett.} \textbf{10}, 045501, (2013).

\bibitem{Ruprecht95}
P. A. Ruprecht, M. J. Holland, K. Burnett, and M. Edwards, \textit{Phys. Rev. A} \textbf{51}, 4704 (1995).

\bibitem{Seman11}
J. A. Seman, R. F. Shiozaki, F. J. Poveda-Cuevas, E. A. L. Henn, K. M. F. Magalh\~{a}es, G. Roati, G. D. Telles and V. S. Bagnato, \textit{ J. Phys.: Conf. Ser.} \textbf{264}, 012004 (2011).


\end{thebibliography}
\end{document}